\documentclass[prl,twocolumn,nofootinbib,10pt,superscriptaddress,preprintnumbers]{revtex4}

\usepackage{graphicx,amssymb,amsmath}

\newcommand{\beq}{\begin{equation}}
\newcommand{\eeq}{\end{equation}}

\newcommand{\abs}[1]{\left\lvert#1\right\rvert}

\newcommand{\cO}{\mathcal{O}}
\newcommand{\cL}{\mathcal{L}}

\newcommand{\cB}{\mathcal{B}}

\newcommand{\Npp}{N^{++}}
\newcommand{\Nmm}{N^{--}}
\newcommand{\Npm}{N^{+-}}
\newcommand{\Nmp}{N^{-+}}
\newcommand{\BB}{B_q-\overline{B}_q}
\newcommand{\BBd}{B_d-\overline{B}_d}
\newcommand{\BBs}{B_s-\overline{B}_s}

\newcommand{\sigtt}{\sigma_{t\bar{t}}}
\newcommand{\ttbar}{t\bar{t}}

\def\mysection#1{{{\bf #1}.~}}

\begin{document}


\preprint{\tiny{CERN-PH-TH/2012-352, CP3-12-56}}

\title{\vspace*{.6cm}Top $B$ Physics at the LHC}

\author{Oram Gedalia}
\email{oram.gedalia@weizmann.ac.il} \affiliation{Department of
Particle Physics \& Astrophysics, Weizmann Institute of
Science, Rehovot 76100, Israel}
\author{Gino Isidori}
\email{gino.isidori@lnf.infn.it} \affiliation{CERN, Theory
Division, CH1211 Geneva 23, Switzerland} \affiliation{INFN,
Laboratori Nazionali di Frascati, Via E. Fermi 40, 00044
Frascati, Italy}
\author{Fabio Maltoni}
\email{fabio.maltoni@uclouvain.be} \affiliation{Centre for
Cosmology, Particle Physics and Phenomenology CP3, Universit\'e
Catholique de Louvain, Chemin du Cyclotron, 1348 Louvain la
Neuve, Belgium}
\author{Gilad Perez}
\email{gilad.perez@cern.ch} \affiliation{Department of Particle
Physics \& Astrophysics, Weizmann Institute of Science, Rehovot
76100, Israel} \affiliation{CERN, Theory Division, CH1211
Geneva 23, Switzerland}
\author{Michele Selvaggi}
\email{michele.selvaggi@uclouvain.be} \affiliation{Centre for
Cosmology, Particle Physics and Phenomenology CP3, Universit\'e
Catholique de Louvain, Chemin du Cyclotron, 1348 Louvain la
Neuve, Belgium}
\author{Yotam Soreq}
\email{yotam.soreq@weizmann.ac.il} \affiliation{Department of
Particle Physics \& Astrophysics, Weizmann Institute of
Science, Rehovot 76100, Israel}

\begin{abstract}
In top-pair events where at least one of the tops decays
semi-leptonically, the identification of the lepton charge
allows to tag not only the top charge but also that of the
subsequent  $b$ quark. In cases where the $b$ also decays
semi-leptonically, the charge of the two leptons can be used to
probe CP violation in heavy flavor mixing and decays. This
strategy to measure CP violation is independent of those
adopted so far in experiments, and can already constrain non
Standard Model sources of CP violation with current and near
future LHC data. To demonstrate the potential of this method we
construct two CP asymmetries based on same-sign and
opposite-sign leptons and estimate their sensitivities. This
proposal opens a new window for doing precision measurements of
CP violation in $b$ and $c$ quark physics via high~$p_T$
processes at ATLAS and CMS.
\end{abstract}

\maketitle

\mysection{Introduction}
The copious production of top quarks at the LHC is usually
exploited to explore various top properties or search for new
heavy resonances. However, it also opens up the possibility to
perform flavor precision measurements. Here we suggest to use
the top quark decay products in order to probe CP violation (CPV) in
heavy flavor mixing and decays.

All existing analyses of CPV in $b$-physics rely on a coherent
production of $b\bar b$ pairs, either from the decay of a $b\bar
b$ resonance or from gluon splitting, where the total $b$
flavor charge at production vanishes. However, top physics
gives another source of $b$'s, and due to the large top mass
and small width, to a good approximation, a top decay yields a
definite non-zero $b$ flavor charge. This charge can be
unambiguously tagged at the time of decay by the charge of the
lepton daughter of the $W$ (originating from the top). In cases
where the $b$ also decays semi-leptonically, we can construct
two CP asymmetries, one in which the latter lepton and the one
from the $W$ are of the same sign, and the other with opposite
signs. In principle, with a good mass resolution one can also
use hadronic decay modes of the $b$; however, this would be
hard  to achieve in the near future at ATLAS and CMS.

To make our discussion more concrete, let us consider  the
interesting result obtained by the D0 collaboration at the
Tevatron on the CP-violating like-sign dimuon
asymmetry~\cite{Abazov:2010hv}:
\begin{align}  \label{eq:AbslD0}
	A^b_{\rm sl}\left({\rm D0} \right)
=	\left( -7.87\pm1.96 \right)\times 10^{-3} \, ,
\end{align}
which differs by 3.8$\sigma$ from the Standard Model (SM)
prediction, $A^b_{\rm sl}\left({\rm SM} \right)
=\left(-3.96^{+0.15}_{-0.04} \right)\times
10^{-4}$~\cite{Lenz:2010gu}. The asymmetries we propose are
conceptually similar although completely independent from
$A^b_{\rm sl}\,$. Similarly, to $A^b_{\rm sl}\,$, our
top-induced CP asymmetries are sensitive to CPV in $\BB$ mixing
($q=d,s$) and to possible exotic sources of direct CPV in $b$
and $c$ decays~\cite{DescotesGenon:2012kr}. As we will show,
these sources appear in different combinations in the two
top-induced CP asymmetries, providing a tool to test the origin
of the anomalous result in Eq.~\eqref{eq:AbslD0}.

Going back to top physics, one can identify three classes of
inclusive top decay chains which produce two leptons of the
same sign:
\begin{align}
&	t \to  \ell^+\nu\, \left(b\to \bar{b}\right) \to\ell^+\,\ell^+\, X \,  , \label{eq:ssb}\\
&	t \to  \ell^+\nu\, \left(b\to c \right) \to\ell^+\,\ell^+\, X \,  , \label{eq:ssc}\\
&	t \to    \ell^+\nu \, \left(b\to \bar{b}\to c \, \bar{c} \right)\to\ell^+\,\ell^+\, X\, , \label{eq:ssccbar}
\end{align}
where throughout this paper $\ell=e,\mu$ and in the process of
Eq.~\eqref{eq:ssccbar} the second $\ell^+$ comes from the $c$
quark and the $\bar{c}$ decays hadronically. These processes
are sensitive to CPV in $\BB$ mixing, semi-leptonic $b$ and $c$
decays and $b\to c$. Similarly, two opposite-sign leptons
emerge from the following processes:
\begin{align}
&	t \to  \ell^+\nu\, b \to\ell^+\,\ell^-\, X \, , \label{eq:osbl}\\
&	t \to  \ell^+\nu\, \left( b \to \bar{b}\to \bar{c} \right) \to\ell^+\,\ell^-\, X \, , \label{eq:osbmix}\\
&	t \to  \ell^+\nu\, \left( b \to c \, \bar{c} \right) \to\ell^+\,\ell^-\, X \, , \label{eq:osccbar}
\end{align}
where in the last process the $\ell^-$ originates from the
$\bar{c}$ quark. Additional negligible contributions via charm
mixing were omitted. We also assume that light mesons can
be rejected by the experimental analysis. 

\mysection{The CP Asymmetries}
The following CP asymmetries related to $\BB$ mixing are
defined:
\begin{align}
	A^{b \ell}_{\rm mix}
=&	\frac{\Gamma\left(b \to \bar b \to \ell^+ X \right) - \Gamma\left(\bar b \to b \to \ell^- X\right)}
	{\Gamma\left(b \to \bar b \to \ell^+ X \right) + \Gamma\left(\bar b \to b \to \ell^- X\right) } \, , \label{eq:Ablmixing} \\
	A^{b c}_{\rm mix}
=&	\frac{\Gamma\left(b \to \bar b \to \bar c~X \right) - \Gamma\left(\bar b \to b \to  c ~X \right)}
	{\Gamma\left(b \to \bar b \to \bar c~X \right) + \Gamma\left(\bar b \to b \to  c ~X \right) } \, .
	\label{eq:Amixing2}
\end{align}
In addition, we define the following direct CPV asymmetries in
the different $b$ and $c$ decay modes:
\begin{align}
	A^{b\ell}_{\rm dir} \label{eq:Abdir}
=&	\frac{ \Gamma\left(b\to\ell^-X \right) - \Gamma\left(\bar b\to\ell^+X \right) }
	{ \Gamma\left(b\to\ell^-X \right) + \Gamma\left(\bar b\to\ell^+X \right)  }\, , \\
	A^{c\ell}_{\rm dir} \label{eq:Acdir}
=&	\frac{ \Gamma\left(\bar c\to\ell^-X_{\rm L}  \right) - \Gamma\left( c\to\ell^+X_{\rm L} \right) }
	{ \Gamma\left(\bar c \to\ell^-X_{\rm L} \right) + \Gamma\left( c\to\ell^+X_{\rm L} \right)  }\, , \\
		A^{bc}_{\rm dir} \label{eq:Abcdir}
=&	 \frac{\Gamma\left( b\to c ~ X_{\rm L}  \right) - \Gamma\left(\bar{b}\to\bar c ~ X_{\rm L} \right)}
	{ \Gamma\left( b\to c ~ X_{\rm L} \right) + \Gamma\left(\bar{b}\to\bar c~ X_{\rm L}  \right)}\, ,
\end{align}
where $X$ ($X_{\rm L}$) denotes an inclusive hadronic final
state  with no leptons and with both light and charm quarks
(with light quarks only). We assume for simplicity no direct
CPV in $b\to c\, \bar{c}$. It is straightforward to generalize
the analysis and incorporate this contribution.

Using these definitions, the same-sign lepton asymmetry in
$t\bar t$ events, $A^{ss}_{\rm sl}\,$, can be decomposed as
follows
\begin{align} \label{eq:Ass}
	A^{ss}_{\rm sl} \equiv
	\frac{\Npp-\Nmm}{\Npp + \Nmm}
=& r_b \, A^{b\ell}_{\rm mix} + r_c \left( A^{bc}_{\rm dir}- A^{c\ell}_{\rm dir} \right) \nonumber\\
&+ r_{c\bar{c}} \left( A^{bc}_{\rm mix} - A^{c\ell}_{\rm dir}\right)  \, ,
\end{align}
with $N^{\pm\pm}$ being the number of events where the sign of
the lepton that originates from the $W$ and the sign of the
lepton from the $b$ are both $\pm$. In addition, we have
defined
\begin{align} \label{eq:rdef}
	r_{q} \equiv
	\frac{\Npp_q+\Nmm_q}{\Npp + \Nmm} \, ,
\end{align}
with $q=b,c,c\bar{c}$ and $N^{\pm\pm}_{b,c,c\bar{c}}$ are the
corresponding numbers of events coming from
Eqs.~\eqref{eq:ssb},~\eqref{eq:ssc} and~\eqref{eq:ssccbar},
respectively, similar to $N^{\pm\pm}$. The $r_q$'s depend on
the choice of the final event selection, designed to enhance
the signal.

Proceeding in a similar way, the opposite-sign lepton asymmetry
in $t\bar t$  events, $A^{os}_{\rm sl}\,$, is defined and
decomposed as follows
\begin{align} \label{eq:Aos}
	A^{os}_{\rm sl}\equiv
	\frac{\Npm - \Nmp}{ \Npm + \Nmp }
=	&\,\tilde{r}_{b} A^{b\ell}_{\rm dir} + \tilde{r}_{c}
\left( A^{bc}_{\rm mix} + A^{c\ell}_{\rm dir}  \right)  \nonumber\\
	&\!\!+\tilde{r}_{c\bar{c}}  A^{c\ell}_{\rm dir} \, ,
\end{align}
where $\tilde{r}_b$, $\tilde{r}_{c}$,  and
$\tilde{r}_{c\bar{c}}$ are the corresponding fractions of
events for the decay chains defined in
Eqs.~\eqref{eq:osbl},~\eqref{eq:osbmix} and~\eqref{eq:osccbar},
respectively (the parameters of the opposite-sign sample are
marked with a tilde).

By construction, all the asymmetries in
Eqs.~(\ref{eq:Ablmixing})--(\ref{eq:Abcdir}) are
phase-convention independent. The mixing asymmetries can be
non-zero either because of CPV in mixing or because of direct
CPV in the subsequent decays of the neutral $B_{s,d}\,$. On the
other hand, the asymmetries in
Eqs.~(\ref{eq:Abdir})--(\ref{eq:Abcdir}) are manifestly due to
direct CPV only. The latter are inclusive partonic asymmetries
that should be interpreted as appropriate averages of the
corresponding exclusive asymmetries involved in a given decay
chain. In principle, the different hadron compositions in
processes with or without mixing (where only the neutral
$B_{s,d}$ mesons are involved) may lead to differences between
the direct CPV asymmetries appearing in $A^{ss}_{\rm sl}$ and
$A^{os}_{\rm sl}\,$. For simplicity, we neglect such
differences.

The expressions of the asymmetries are greatly simplified in
the limit where we can neglect direct CPV. In this limit
$A^{q\ell}_{\rm dir} = A^{bc}_{\rm dir} =0$, and the mixing
asymmetries can be related to the theoretical parameters
describing meson-antimeson mixing. Following the convention
of~\cite{Grossman:2009mn} we have
\begin{align}
	A^{b\ell}_{\rm mix}
&= A^{bc}_{\rm mix} =
f_d a^d_{\rm SL} + f_s a^s_{\rm SL} \nonumber \\
&= 	f_d\frac{1-\abs{q_{B_d}/p_{B_d}}^4}{ 1+\abs{q_{B_d}/p_{B_d}}^4}
	+f_s\frac{1-\abs{q_{B_s}/p_{B_s}}^4}{1+\abs{q_{B_s}/p_{B_s}}^4 } \, ,
\end{align}
where $q_X$ and $p_X$ are the parameters describing the mass
eigenstates in the flavor basis and $f_{d,s}$ are the fractions
of $b$ quarks forming $B_{d,s}$ mesons.

\mysection{LHC Sensitivity}
The sensitivity of the proposed measurements can be naively
estimated by counting the expected number of events and
deriving the statistical uncertainty. Systematic uncertainties
are not taken into account here. We consider only the dominant
production mechanism, namely of top-pairs. In principle, the
contribution of single tops can be incorporated by using an
appropriate data-based normalization to compensate for the
different production rates of tops and anti-tops at the LHC.
Yet, the statistical gain is small; hence we do not include
such a signal in our analysis.

We focus on events where one of the tops decays
semi-leptonically. The resulting lepton enables to tag the
charges of the $b$ quarks from the top and the anti-top, such
that both can be included in the analysis. The association of
each $b$ jet ($b$-charge association) with the appropriate top
is done by using the matrix element method, as discussed below.
Note that events where both $b$ and $c$ from the same top decay
semi-leptonically are rejected. In principle, one could extend
the analysis to include such finals states; however, their
inclusion makes the analysis more complicated without a
significant gain in sensitivity.

We use Monte-Carlo tools to study the efficiencies of the
$b$-charge association and  the kinematical cuts. The $t \bar t$ sample of events at $\sqrt{s} = 14$~TeV is generated using
MadGraph/MadEvent 5 v1.5.5~\cite{Alwall:2011uj}, Pyhtia~6.4~\cite{Sjostrand:2006za} and
DELPHES~2.0.3~\cite{Ovyn:2009tx} for the detector response. 
In order to  capture QCD radiation effects, we 
have included $t \bar t$ events with up to three extra partons employing the MLM-$k_T$ merging procedure~\cite{Alwall:2008qv}. 
We select events with at least one charged lepton ($p_T >  10$~GeV)
and four jets ($p_T >  20$~GeV), two of which are $b$-tagged. It is interesting to note that the requirement of two $b$ jets
ensures that a potential contribution of CPV in $t\to b$ decays
is absent in the sample.

The number of events in each channel is given by
\begin{align} \label{eq:Nq}
    N^{\pm\pm}_q \left( N^{\pm\mp}_q \right)
=   \sigtt \, \cL \, {\rm BR}\!\left(t \bar{t} \to b\bar{b}\ell \nu\, {\rm had} \right) \,
    \epsilon_{\rm sel}\epsilon_b^2\epsilon_{\rm A} \,\cB_{q}\,,
\end{align}
where $q$ represents the various processes in
Eqs.~\eqref{eq:ssb}--\eqref{eq:osccbar}, $\sigtt$ is the
top-pair production cross section, $\cL$ is the integrated
luminosity, ${\rm BR}\!\left(t \bar{t} \to b\bar{b} \ell \nu\,
{\rm had}\right)\approx0.30$, $\epsilon_{\rm sel}\approx0.55$
is the efficiency of selecting the lepton and the four jets,
$\epsilon_b\approx0.60$ is the $b$-tagging efficiency and
$\epsilon_{\rm A}\approx0.70$ is the $b$-charge association
efficiency (see below). In addition, for each of the processes
in Eqs.~\eqref{eq:ssb}--\eqref{eq:osccbar} we have
\begin{align}
	\cB_b \label{eq:cBb}
=&	 {\rm BR}\!\left(b \to \ell\right) \overline{\chi}\left[1-{\rm BR}\!\left( b \to c \to\ell\right) \right]
=	0.024 \, , \\
	\cB_c \label{eq:cBc}
=&	  {\rm BR}\!\left(b\to c\to \ell \right)\left[1-{\rm BR}\!\left( b \to\ell\right) \right]
=	0.12 \, , \\
	\cB_{c\bar{c}} \label{eq:cBcc1}
=&	  {\rm BR}\!\left(b\to \bar{c}\to \ell \right) \overline{\chi} \nonumber\\
&\times\left[1-{\rm BR}\!\left( b \to c \to\ell\right) \right]
=	3.4\times10^{-3} \, , \\
	\tilde{\cB}_{b} \label{eq:cBbt}
=&	 {\rm BR}\!\left(b \to\ell\right) \left(1-\overline{\chi}\right) \nonumber \\
&\times \left[1-{\rm BR}\!\left( b \to c\to\ell\right)\right]
=	0.17 \, , \\
	\tilde{\cB}_{c} \label{eq:cBc1t}
=&	 {\rm BR}\!\left(b\to c\to \ell \right)\overline{\chi} \left[1-{\rm BR}\!\left( b \to \ell\right) \right]
=	0.016 \, , \\
&\times\left[1-{\rm BR}\!\left( b \to \ell\right)\right]
=	4.7\times 10^{-4} \, ,\\
	\tilde{\cB}_{c\bar{c}} \label{eq:cBcct}
=&	  {\rm BR}\!\left(b\to \bar{c}\to \ell \right) \nonumber\\
&\times\left[1-{\rm BR}\!\left( b \to c \to\ell\right) \right]
=	 0.027\, ,
\end{align}
respectively. Here ${\rm BR}\!\left(b \to \ell\right)=0.23$
(for $e$ and $\mu$ including leptonic $\tau$'s), ${\rm
BR}\!\left(b\to c\to \ell \right)=0.16$ and ${\rm
BR}\!\left(b\to \bar{c}\to \ell \right)=0.032$, where the last
two are  without $B$ mixing~\cite{Nakamura:2010zzi}.
Furthermore, $\overline{\chi}=0.13$ is the mixing probability
for a $b$~quark~\cite{Amhis:2012bh}. The last factor in each of
the above equations removes events where both $b$ and $c$ (or
$c$ and $\bar{c}$ in $b\to c\,\bar{c}$ events) decay
semi-leptonically (assuming that $\mathrm{BR}\left(b \to
c\right) \approx1$). The $r_q$'s can be calculated from
Eqs.~\eqref{eq:Nq}--\eqref{eq:cBcct}
\begin{align} \label{eq:rqvalues}
	r_b &= 0.16 \, , \quad
	r_c = 0.82 \, ,  \quad
	r_{c\bar{c}} = 0.022 \, , 	\nonumber \\
	\tilde{r}_{b} &= 0.79 \, \quad,
	\tilde{r}_c = 0.075 \, , \quad
	\tilde{r}_{c\bar{c}} = 0.13	\, .
\end{align}

The statistical uncertainty in estimating the asymmetries is
given by $\delta A_{\rm sl}^{ss}=1/\sqrt{\Npp+\Nmm}$ and
similarly for the opposite-sign. Plugging in the above numbers
leads to
\begin{align}
	\delta A^{ss}_{\rm sl}=\frac{9.0}{\sqrt{ \sigtt\, \cL}} \,, \qquad
    \delta A^{os}_{\rm sl}=\frac{7.6}{\sqrt{ \sigtt\, \cL}} \,.
	\label{eq:delA}
\end{align}

\begin{figure}[tb]
  \centering
  \includegraphics[width=.45\textwidth]{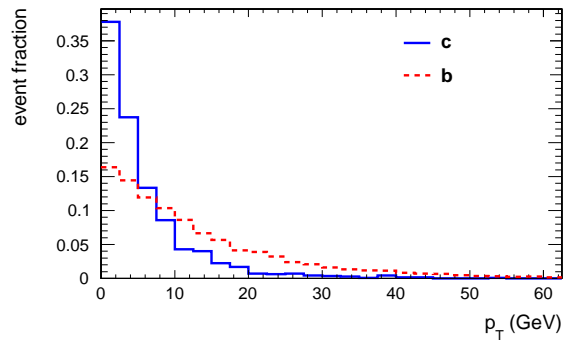}
  \caption{The $p_T$ distribution of the lepton coming from the $b$ quark (dotted red)
  and the one coming from the $c$ quark (solid blue). The distributions
  are normalized separately.}
  \label{fig:ptdist}
\end{figure}

The measured asymmetries $A_{\rm sl}^{ss,os}$ can be used to
extract information on the various CPV sources in
Eqs.~\eqref{eq:Ablmixing}--\eqref{eq:Abcdir}. One may hope that
the sensitivity to each of these sources separately will be
improved by applying appropriate kinematical cuts (thus
changing the values in Eq.~\eqref{eq:rqvalues}). In particular,
it is expected that the lepton coming from a $b$ semi-leptonic
decay would be more energetic than the lepton from a subsequent
$c$ decay. The corresponding $p_T$ distributions are plotted in
Fig.~\ref{fig:ptdist}. A detailed analysis of the selection
criteria may lead to an improved sensitivity, but in our study
we found no significant gain.

If we neglect the direct CPV sources, the measurement of
$A_{\rm sl}^{ss,os}$ can be used to estimate the CPV in $\BB$
mixing:
\begin{align}
\delta A^{b\ell}_{\rm mix}
\approx 7\times 10^{-3}\left(3\times 10^{-3} \right), \ \
\end{align}
for $\sqrt{s}=14$~TeV and $\cL=50\left(300\right)$~fb$^{-1}$
using $\sigtt=852$~pb~\cite{Czakon:2012zr}. This sensitivity is somewhat inferior to the result of
Eq.~\eqref{eq:AbslD0}. The $B$-factories report the flavor-specific CP asymmetry in $\BBd$ with a combined sensitivity of $3\times10^{-3}$~\cite{Amhis:2012bh}. Furthermore, the sensitivity obtained by LHCb for $\BBs$ mixing is $6\times10^{-3}$, and it is expected to improve to the permil level by 2018~\cite{Bediaga:2012py}.

Similarly, our proposed measurements will be able to provide
strong upper bounds on the direct CPV sources in
Eqs.~\eqref{eq:Abdir}--\eqref{eq:Abcdir} in case of null
signals. With 50~fb$^{-1}$ at 14~TeV we can obtain
\begin{align}
	\abs{ A^{b\ell}_{\rm dir}} \lesssim 0.3\% \, , \
	\abs{ A^{c\ell}_{\rm dir}} \lesssim 0.3\% \, , \
	\abs{ A^{bc}_{\rm dir}} \lesssim 0.3\% \, ,
\end{align}
at 2$\sigma$, assuming no cancellations. The first two bounds
above are stronger than existing bounds of 1.2\% and 6\%,
respectively~\cite{DescotesGenon:2012kr}. Even with the 8~TeV
run, bounds of 1\% or better can be obtained for the direct CPV
sources.

In the above discussion we assumed prefect identification of
the lepton originating from the $B$ meson. We now consider the
systematic uncertainty induced by a wrong association of the
$b$ with the top. The observed number of events $N^{XY}_*$ is
then given by
\begin{align}
	N^{\pm\pm}_* =\left(1-\epsilon_{\rm F}\right) N^{\pm\pm} + \epsilon_{\rm F} N^{\mp\pm} \, , \\
	N^{\pm\mp}_* =\left(1-\epsilon_{\rm F}\right) N^{\pm\mp} + \epsilon_{\rm F} N^{\mp\mp} \, ,
\end{align}
where $\epsilon_{\rm F}$ is the probability for a wrong
association. The measured CP asymmetries, $A^{ss,os}_{{\rm
sl}\,*}$, can be related to $A^{ss,os}_{{\rm sl}}$ as follows
\begin{align}
	A^{ss}_{{\rm sl}\,*} \label{eq:AssS}
\approx A^{ss}_{\rm sl} - \epsilon_{\rm F} \frac{\Npm}{\Npp}\left( A^{os}_{\rm sl} + A^{ss}_{\rm sl} \right) \, , \\
	A^{os}_{{\rm sl}\,*} \label{eq:AosS}
\approx A^{os}_{\rm sl} - \epsilon_{\rm F} \frac{\Npp}{\Npm}\left( A^{ss}_{\rm sl} + A^{os}_{\rm sl} \right) \, ,
\end{align}
where we expand to first order in $\epsilon_{\rm F}$ and in the
asymmetries and $\Npm/\Npp\sim 1.4$. We learn that as long as
$\epsilon_{\rm F}\lesssim 10\%$, the error in the measured mean
values would be smaller than the estimated statistical
uncertainty. If $\epsilon_{\rm F}$ is known to a good accuracy,
then the two equations above can be used to extract
$A^{ss,os}_{\rm sl}$ from the measured asymmetries.

\mysection{$b$-charge Association}
Given a pure semi-leptonic $\ttbar$ sample with two
reconstructed $b$ jets, $b_1$ and $b_2$, two light jets and at
least one charged lepton, our goal is to determine the charge
of $b_1$ and $b_2$. We call this procedure $b$-charge
association. Naively, this can be done by reconstructing the
top mass. Without any particular optimization, this method
gives a high misassociation rate ($\epsilon_{\rm
F}\approx35\%$). Hereafter we propose an alternative method.

The matrix element method has been successfully used in the
determination of the top quark mass~\cite{Abazov:2006bd} and
the first single top observation at the
Tevatron~\cite{Aaltonen:2009jj}. One can compute the
probability that a given experimental event originates from
some process
\beq
P(x) = \frac{1}{\sigma}\int dy|M(y)|^2T(x|y) \,,
\label{eqn:pbar}
\eeq
where $\sigma$ denotes the effective cross section, $M$ is the
partonic amplitude and $T$ is the transfer function, which
gives the probability of reconstructing particles of momenta
$x$ originating from parton level momenta $y$.

\begin{figure}[tb]
  \centering
  \includegraphics[width=.45\textwidth]{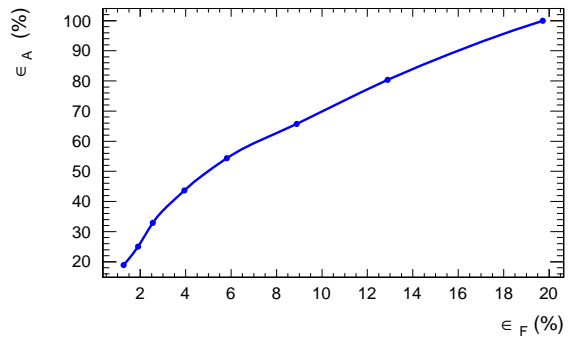}
  \caption{Efficiency of the $b$-charge association as a function of the misassociation probability.}
  \label{fig:effmis}
\end{figure}

We use MadWeight~\cite{Artoisenet:2010cn} to compute two
probabilities $P_1$ and $P_2$ per event, corresponding to the
two possible associations of $b_1$ and $b_2$ with the initial
partons $b$ and $\bar{b}$. The larger probability tells us
which configuration is more likely. The larger the difference
between $P_1$ and $P_2\,$, the more confident we are in the
association. The discriminant variable
$W\equiv(P_1-P_2)/(P_1+P_2)$ can be used to interpolate
continuously between a low purity (high efficiency) and high
purity (low efficiency) $b$-association. In
Fig.~\ref{fig:effmis} the efficiency vs.\ the misassociation
rate is shown. The working point is chosen such that
$\epsilon_{\rm F}\approx$~10\%, which corresponds to a signal
efficiency $\epsilon_{\rm A}\approx$~70\%. We find that the impact of removing the matching slightly increases $\epsilon_{\rm A}$ by order of 5\%. 

\mysection{Relation to the D0 Dimuon Anomaly}
The possibility to explain the D0 anomaly by allowing not only
CPV in $\BB$ mixing but also direct CPV in $b$ and $c$
semi-leptonic decays was discussed
in~\cite{DescotesGenon:2012kr}. This was done by considering
each of these CPV sources separately while taking the other two
to be SM-like. It was found that the D0 result can be
accommodated by the value in Eq.~\eqref{eq:AbslD0} for $\BB$
mixing, $A^{b\ell}_{\rm dir}\left({\rm D0}
\right)\sim(3\pm1)\times 10^{-3}$ or $A^{c\ell}_{\rm
dir}\left({\rm D0} \right)\sim(9\pm3)\times 10^{-3}$. Within
the SM, the latter two are $A^{b\ell}_{\rm dir}\left( {\rm SM}
\right)\sim10^{-7}$ and $A^{c\ell}_{\rm dir}\left( {\rm SM}
\right)\sim10^{-11}$~\cite{DescotesGenon:2012kr}.

For each of these cases, we can estimate the asymmetries
$A^{ss,os}_{\rm sl}$ and the discrimination power in measuring
them (assuming no CPV in $b \to c$ decays). If
the D0 result originates from direct CPV in semi-leptonic charm
decays, $A^{ss}_{\rm sl}$ should be non-vanishing at
significance of 2.8$\sigma$ with 50~fb$^{-1}$ at 14~TeV.
Similarly, $A^{ss}_{\rm sl}$ ($A^{os}_{\rm sl}$) should be
probed at 2.1$\sigma$ (2.9$\sigma$) for CPV in $\BB$ mixing
(direct CPV in semi-leptonic $b$ decays) for the 14~TeV run
with a sample of 300~fb$^{-1}$.

\mysection{Discussion}
In this letter we have proposed to probe CP violation in $B$
mixing and in $b$ and $c$ decays in top-pair events, by
exploiting the $b$-charge tagging ability inherent to the top
(semi-leptonic) decay. This presents a striking opportunity to
explore low-energy flavor observables in processes at a much
higher scale of $\cO\!\left(100\right)$~GeV.

Given the estimated uncertainties, a significant non-zero
signal in each of the asymmetries introduced above will
unambiguously imply the existence of new physics beyond the SM
(which gives $A^{ss,os}_{\rm sl}({ \rm SM})<10^{-4}$). The
sources of the systematic uncertainties in this measurement
are different than those of other experiments such as LHCb and
the $B$-factories, hence it will serve as an important
contribution to the study of CP violation in the quark sector,
even if the overall sensitivity is lower.

There are a few issues in the above analysis which call for a
more detailed study. First, we have neglected systematic
effects at the detector level that might lead to an asymmetry
in the measured rates of leptons vs.\ anti-leptons. Second, we have only partially included higher-order QCD effects (up to three extra matched jets) in our $t\bar{t}$ sample simulation, and not the full NLO contributions. At the LHC these effects are known to induce a (small) charge asymmetry
between the rapidity distributions of
the top and the anti-top. Depending on the details of the
selection and the analysis, these effects might feed down to
the asymmetries $A^{ss,os}$. The impact of such detector and
physics effects can be studied in data.

Concerning the backgrounds, we have verified that our selection
and reconstruction procedure keeps the contribution of $W+$jets
low enough to be neglected at this proposal stage.  $W+$jets is
charge asymmetric and therefore could alter the asymmetries. In
addition, backgrounds could affect the misassociation rate. It
is reasonable to believe that a significant fraction  of these
processes can be rejected with an appropriate $b$-tagging
algorithm and the $b$-charge association selection.

As further improvement, we note that the sample statistics
could be increased by including $t \bar t$ events where only
one $b$ jet is tagged. In that case, it could be advantageous
to incorporate single top events into one combined analysis,
thus avoiding  the need to treat them as background, by
suitably redefining the asymmetries $A^{ss,os}$ in terms of
fractions of events  relative to the measured (and different)
top and anti-top rates.

Last but not least, we mention that this study could be
extended by analyzing the time-dependence of $B$ decays
produced from top decays. Indeed the lepton from the initial
$W$ does not only provide a perfect flavor tag, it also
provides an indication of the position (time) where (when) the
flavor eigenstate has been produced. This allows us to
reproduce in high~$p_T$ physics a time- and flavor-tag
configuration conceptually similar to that obtained at the
$B$-factories.

\mysection{Acknowledgments}
We thank Alex Cerri and Yosef Nir for discussions; FM and GP
thank the Corfu Summer School 2012 where the seeds of this
project were planted. GI acknowledges the support of the EU ERC
Advanced Grant FLAVOUR (267104). FM and MS are supported by the
IAP Program, BELSPO VII/37. GP is the Shlomo and Michla Tomarin
chair, supported by the grants from Gruber foundation, IRG, ISF
and Minerva.



\end{document}